\documentclass[
  aps,
  prl,
  reprint,
  longbibliography
]{revtex4-2}

\usepackage[T1]{fontenc}
\usepackage[utf8]{inputenc}
\usepackage{lmodern}
\usepackage{graphicx}
\usepackage{braket}
\usepackage{physics}
\usepackage[caption=false]{subfig}
\usepackage[colorlinks=true,linkcolor=black,citecolor=black,urlcolor=black]{hyperref}
\usepackage{microtype}
\usepackage{exscale}
\usepackage{relsize}

\newcommand{\subfig}[2]{%
	{\textsf{#1}} \vtop{%
		\vskip0pt
		\hbox{#2}
	}}

\begin{document}

    \title{Tuning the universality class of a quantum process by Trotterization}
    \author{Robin Prampain dit Boulan}
    \email {r.prampain.dit.boulan@tu-berlin.de}
    \author{Jerom Jebai}
    \author{Hendrik Weimer}
    \affiliation{Institut f\"ur Physik und Astronomie, Technische Universit\"at  Berlin, Hardenbergstraße 36, EW 7-1, 10623 Berlin, Germany}
    \date{\today}

\begin{abstract}
We numerically analyze a discretized version of the quantum contact
process, a non-equilibrium model having a competition between a
coherent infection process and a dissipative recovery
process. Previously, the continuous-time version has been discussed to
have a phase transition in a novel quantum universality class, while
experimental investigations of a related process have pointed towards
the transition being in the classical directed percolation class. We
demonstrate that these differences stem from details of the
discretization of the process and construct a systematic way to
continuously interpolate between these results. Finally, we provide
evidence that the modification of the universality class is due to
destructive interference effects, showing a clear quantum-mechanical
origin.
\end{abstract}

\maketitle

\begin{figure}[t]
  \begin{tabular}{ll}
    \subfig{a}{\includegraphics[height=3cm]{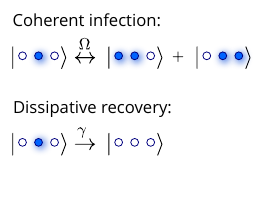}}&\subfig{b}{\includegraphics[height=3cm]{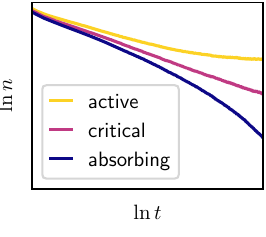}}
  \end{tabular}
  
  \subfig{c}{\includegraphics[width=0.9\linewidth]{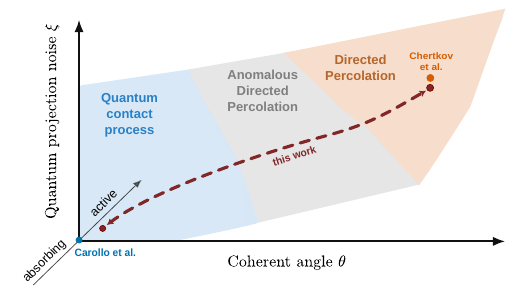}}
  \caption{Quantum contact process. (a) Basic processes consisting of coherent infection and dissipative recovery. (b) Characteristic behavior of the density of infected cells $n(t)$ near the critical point. (c) Sketch of the different universality classes at the critical point of the Trotterized quantum contact process. Depending on the Trotter angle $\theta$ and the strength of the projection noise $\xi$, three different regions are found, corresponding to directed percolation, a novel quantum contact process universality class connected to the continuous-time process, and an intermediate region exhibiting anomalous directed percolation. The path through this parameter space investigated here is shown as a dashed line, while the prior works on the continuous-time case and the experimentally studied Floquet quantum contact process are shown as individual points.}
    \label{fig:intro_sketch}
\end{figure}
Open quantum many-body systems are a fascinating platform to explore
phenomena beyond closed quantum systems \cite{Kasprzak2006,Lai2007,Syassen2008,Verstraete2009,Baumann2010,Carusotto2013,Lemeshko2013a,Marcuzzi2016,Letscher2017,Helmrich2020,Wu2024,Mi2024,Wu2024b}, but their coupling
to an external environment invariably brings up the question whether
their behavior is still genuinely quantum-mechanical. In this context,
numerical evidence for the quantum contact process exhibiting a novel
quantum universality class at criticality is a rare case where strong
quantum effects potentially play a dominant role
\cite{Buchhold2017,Carollo2019,Gillman2019,Jo2021}. However, experimental results have hinted towards
the phase transition being in the classical directed percolation class
\cite{Chertkov2023}. Here, we resolve this controversy and demonstrate
that the distinction can be attributed to differences in the
discretization of the process and show that the novel universality
class is closely linked to inherently quantum-mechanical
interference effects.

The classical counterpart of the quantum contact process has been
studied extensively, as it is a paradigmatic model for epidemic
spreading without immunization \cite{Hinrichsen2000}. In both the
classical and quantum case, there is a phase transition between an
absorbing phase where all infections die out and an active phase
exhibiting a finite density of infections in the stationary state, see
Fig.~\ref{fig:intro_sketch}. Due to the non-equilibrium nature of the
process, it is even possible to observe the transition in
one-dimensional systems. Crucially, numerical evidence for the quantum
contact process points to the transition belonging to a different
universality class compared to the classical case of directed
percolation \cite{Carollo2019}. This makes the quantum contact process
a unique candidate to observe quantum criticality with mixed quantum
states \cite{Sieberer2025} and a strong candidate to demonstrate a
quantum advantage in quantum simulation \cite{Weimer2023}.

In the case of the classical contact process, the details of the
discretization are irrelevant in the sense of the renormalization
group. Hence, large-scale Monte-Carlo studies use discrete timesteps
for faster simulations. In the quantum case, however, it is not clear
whether details of the discretization are irrelevant as well
\cite{Makki2024}. This is exacerbated by the fact that there are
different ways to discretize the dynamics, e.g., by using a
Suzuki-Trotter expansion or by randomly placing reset gates into a
circuit representation of the contact process \cite{Chertkov2023}.

In this Article, we introduce a systematic way to continuously deform
the discretization of the quantum contact process. In our work, we
always tune the system to criticality and compute the long-time decay
of the density to determine the universality class using
matrix-product state simulations. Starting near the continuous case
within a Suzuki-Trotter expansion, we increase the stepsize of the
Trotterization until we end in a situation that is close to the
experimental realization using trapped ions, see
Fig.~\ref{fig:intro_sketch}. Strikingly, we find that the increased
stepsize results in an increased quantum projection noise that
dramatically modifies the underlying phase transition. For small
Trotter angles, we observe the universality class associated with the
continuous-time quantum contact process, while for large angles, we
recover directed percolation as in the experiment. In between, we
observe a third region exhibiting a continuous variation of the
critical exponent that is consistent with anomalous directed
percolation. We attribute the differences in the universality class to
the emergence of dark states formed by destructive interference
between different infection pathways, establishing strong evidence
that the modification of the phase transition has a genuinely
quantum-mechanical origin.

\section{Trotterized quantum contact process}

We consider the quantum contact process on a one-dimensional lattice
with the states $\ket{0}$ and $\ket{1}$ describing healthy and
infected cells, respectively. The continuous-time version can be
described by a Markovian quantum master equation in Lindblad form
\begin{equation}
\frac{d}{dt}\rho = -i[H,\rho] + \sum\limits_{i} \left(c_i\rho c_i^\dagger - \frac{1}{2}\left\{c_i^\dagger c_i, \rho\right\}\right).
\label{eq:master}
\end{equation}
The coherent part describes the coherent infection process, according
to the Hamiltonian
\begin{equation}
  H = \Omega \sum\limits_{i=1}^L (n_i \sigma^x_{i+1} + \sigma^x_i n_{i+1}),
\end{equation}
where $n_i = (1-\sigma^z_i)/2$ is the density and $\sigma^\alpha$
refers to the Pauli matrices. The quantum jump operators $c_i$ describe
the dissipative recovery and have the form $c_i=\sqrt{\gamma}
\sigma^-_i = \sqrt{\gamma} (\sigma^x_i + i\sigma^y_i)/2$. The
resulting dynamics is then determined by the ratio $\Omega/\gamma$ of
the two coupling constants.

\begin{figure}[t]
  \includegraphics[width=0.95\linewidth]{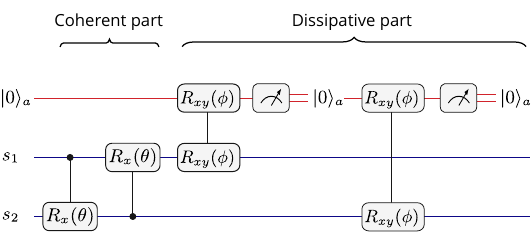}
  \caption{Discrete dynamics of a single timestep for $L=2$. System qubits $s_i$ are coupled to an ancilla qubit $a$ initialized in $\ket{0}$. The coherent dynamics is performed via Controlled-$R_x$ rotation gates on the system qubits with the coherent angle $\theta$. The dissipative step is performed by coupling each system qubit to the corresponding ancilla using a parametric swap ($R_{xy}$) gate with the dissipative angle $\phi$. Finally, the ancilla is reset to $\ket{0}$, providing a dissipative element.}
  \label{fig:gates}
\end{figure}
We perform a discretization of the quantum contact process using two
guiding principles. First, we recover the continuous-time process in
the limit of vanishing discretization parameters. Second, we perform
independent blockwise discretizations of the coherent and dissipative
parts, as this is in line with the experimental Floquet quantum
contact process \cite{Chertkov2023}. Both principles can be understood
in terms of a first-order Suzuki-Trotter expansion of the dynamical
map $\mathcal{V}\rho =\exp(\mathcal{L}\tau)\rho$, with $\mathcal{L}$
being the Liouvillian, according to $\mathcal{V}\rho =
\mathcal{V}_d\mathcal{V}_c\rho + O(\tau^2)$. The coherent part
$\mathcal{V}_c\rho = U(\theta)\rho U^\dagger(\theta)$ and can readily
be decomposed into a circuit form using elementary $R_{xx}$ and $R_{yy}$ quantum gates, see Fig.~2. The dissipative part $\mathcal{V}_d$ can be realized by introducing an ancilla qubit, which is reset to the state $\ket{0}$ at the end of the sequence. The map $\mathcal{V}_d=\prod_i\mathcal{V}_{d,i}$ can be expressed in terms of Kraus operators $K_{i,\alpha}$ as
\begin{equation}
  \mathcal{V}_{d,i}\rho=K_{i,0}(\phi)\,\rho\,K_{i,0}^\dagger(\phi)+K_{i,1}(\phi)\,\rho\,K_{i,1}^\dagger(\phi),
\end{equation}
with $\phi$ being the angle in the swap operation between system and
ancilla, see Fig.~\ref{fig:gates}. During each timestep $\tau$, the
probability to reset the system qubit to $\ket{0}$ is given by
$p=\sin^2(\phi/2)$, which translates to a strength of the quantum
projection noise $\xi = \sin(\phi)/2$, see the Methods section for
details. In the continuous-time limit, the noise per timestep goes to
zero. Crucially, the damping of the off-diagonal coherences of the
matrix also becomes weaker, see the methods section for
details. This preservation of coherences opens the possibility to have
genuine quantum effects driving the phase transition of the quantum
contact process.

We are now in the position to continuously change $\phi$ from a
continuous-time process $\theta,\phi \to 0$ corresponding to
continuous weak measurements of the system, to a process undergoing
strong projective measurements at $\phi=\pi/2$. In the following, we
choose a fixed coherent angle $\theta$ and tune the system into
criticality by varying $\phi$. We note that there are also other
possibilities to modify the strength of infection and recovery,
respectively, e.g., by randomly choosing which sites undergo a
dissipative process during each timestep \cite{Chertkov2023}.

\section{Matrix product state simulations}

\begin{figure*}[t]
  \begin{tabular}{cc}
    \subfig{a}{\includegraphics[width=8.5cm]{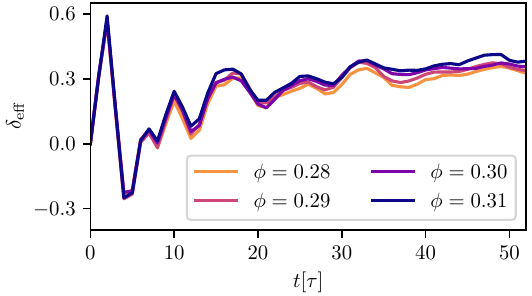}} &\subfig{b}{\includegraphics[width=8.5cm]{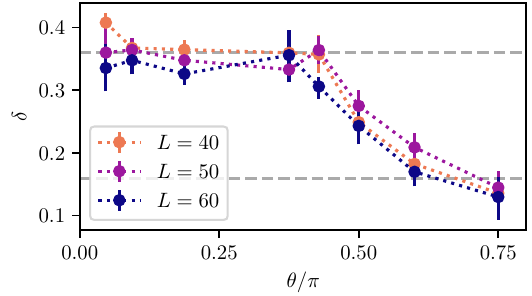}}
  \end{tabular}
  \caption{Effective critical exponent of the Trotterized contact
    process. (a) In the long-time limit, $\delta_\text{eff}(t)$
    approaches the QCP value of 0.36 for a small Trotter angle $\theta
    = 3\pi/32$ ($L=60$). (b) For larger values of $\theta$, the
    critical exponent $\delta$ according to the long-time limit
    continuously drops from the QCP value towards the DP value of
    0.159 (dashed lines). The behavior is consistent for different
    system sizes, ruling out strong finite-size effects. Error bars
    are standard errors from bootstrapping with 100 samples.}
  \label{fig:delta}
\end{figure*}
We analyze the long-time dynamics of the Trotterized contact process
using matrix product state simulations with a system size up to $L=60$
cells, see the Methods section for details. Specifically, we monitor
the average density $n(t) = \sum_i n_i(t)/L$. Then, the long-time
behavior at criticality follows the scaling form
\begin{equation}
  \rho \sim t^{-\delta} \chi([\phi-\phi_c]^{1/\nu_\parallel},t^{1/z}/L),
\end{equation}
where $\delta$, $\nu_\parallel$, and $z$ are critical exponents, $\phi_c$
is the critical value of $\phi$ for fixed $\theta$, and $\chi$ is a
scaling function \cite{Hinrichsen2000}. Numerically, the universality
class can be determined by tuning the system to criticality,
$\phi=\phi_c$, and extracting the $\delta$ exponent according to
\begin{equation}
  \delta_{\text{eff}}(t) = -\frac{1}{\log m}\log\frac{n(mt)}{n(t)}
  \label{eq:delta}
\end{equation}
which approaches a constant value in the long-time limit for a fixed
value of $m$ \cite{Hinrichsen2000}. For the classical directed
percolation (DP) universality class, the critical exponent takes a
value of $\delta = 0.159$, whereas in the quantum contact process
(QCP), $\delta\approx 0.36$ has been reported \cite{Carollo2019}.

For the quantum contact process, there have been observations that the
rapidity-reversal symmetry might be broken \cite{Gillman2019}, i.e.,
the exponent $\delta$ could be different when starting with a fully
infected lattice compared to a single seed infection. To avoid this
source of ambiguity, we always start from a fully infected initial
state.

The results of our simulations are presented in
Fig.~\ref{fig:delta}. For each value of $\theta$, we observe a fast
convergence of $\delta_\text{eff}$ to the value in the long-time limit. Up
to a value of $\theta\approx 1.2$, the critical exponent is close to
the QCP value of $\delta=0.36$. Afterwards, there is a continuous decline,
until it approaches the DP value of $\delta=0.159$ for $\theta \approx 3\pi/4$,
consistent with the experimental results of \cite{Chertkov2023}. We
note that the intermediate regime with a continuously varying critical
exponent is consistent with the transition belonging to the anomalous
directed percolation class \cite{Hinrichsen2000}. However, in contrast
to the recently discussed case of facilitation in Rydberg systems
\cite{Brady2024}, our model does not exhibit long-range interactions
on the microscopic level. Instead, these processes would have to be
generated under the renormalization group flow. The analysis of this
intermediate region is complicated by the aforementioned possibility
of rapidity-reversal symmetry being broken; this precludes standard
tests for anomalous directed percolation based on hyperscaling relations
\cite{Hinrichsen2000}.

Alternatively, a continuously varying exponent $\delta$ can also
emerge in a Griffiths phase, which has previously been discussed in
the context of disordered systems
\cite{Vojta2005,Wintermantel2021,Brady2024b}. While certain clean
quantum systems are known to exhibit effects typically found in
disordered systems \cite{Schiulaz2015}, numerical evidence points
against the existence of a Griffiths phase in our case. Crucially, a
Griffiths phase would exhibit a power-law decay of $n(t)$ over an
extended range of $\phi$ values. This would manifest itself in
$\delta_\text{eff}(t)$ becoming constant in the long-time limit, while
the value of the constant would change with the angle $\phi$. Instead,
we find a constant value of $\delta_\text{eff}(t)$ only around an
isolated critical value of $\phi$, constituting strong evidence
against the existence of a Griffiths phase.

\section{Emergence of dark states}

\begin{figure}[t]
  \centering
  \subfig{a}{\includegraphics[width=0.95\linewidth]{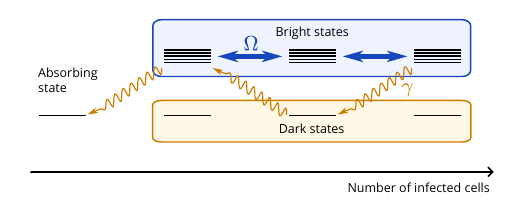}}
  \subfig{b}{\includegraphics[width=0.95\linewidth]{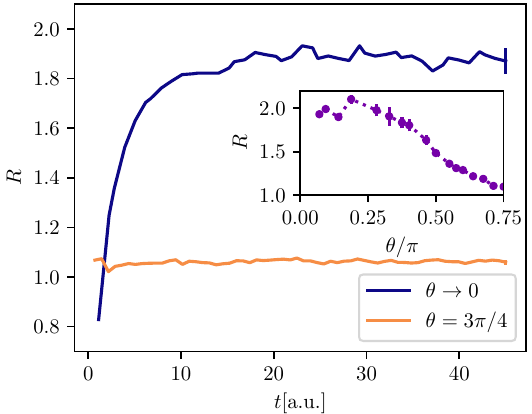}}
    \caption{Emergence of dark states. (a) Strong coherent infection
      coupling $\Omega$ only couples the bright states. Weak
      dissipation $\gamma$ can trap the system in a dark state, in
      which no infections are possible until the next dissipative
      event. (b) Ratio $R(t)$ between dark and bright state pairs at
      criticality, showing a clear difference between the
      continuous-time case belonging to the QCP universality class and
      $\theta=3\pi/4$ belonging to the DP universality class. Error
      bars denote represenative Monte-Carlo sampling errors. The inset
      shows that the drop from $R(t\to\infty)\approx 2$ occurs around
      the $\theta$ value where the universality class starts to
      deviate from QCP.}
       \label{fig:dark}
\end{figure}
While we provide strong evidence for the existence of a distinct
universality class describing the quantum contact process, the
mechanism behind its formation requires further investigation. A key
question is whether the change in the transition can be attributed to
genuine quantum effects or whether the same physical behavior can also
be found in classical systems. In the following, we present evidence
for the change in universality class arising due to the emergence of
dark states and their destructive interference effects, pointing to a
genuine quantum origin.

Dark states refer to the existence of zero-energy eigenstates of the
Hamiltonian, as they do not show any coherent dynamics under the
Schr\"odinger equation. They typically arise from destructive
interference between excitation and de-excitation terms in the
Hamiltonian. Dark states play a key role in quantum phenomena like
electromagnetically-induced transparency \cite{Fleischhauer2005} and
have important applications in a large class of quantum state
preparation protocols
\cite{Diehl2008,Kraus2008,Weimer2010}. In our case, the
initial fully infected state has no overlap with any dark
states of the Hamiltonian, which are thus not accessible from coherent
dynamics alone. However, dissipation can lead to a decay into a dark
state of the Hamiltonian, where the system remains trapped until a
subsequent decay event can again produce a bright state (i.e., a state
which is not a dark state).

Note that the suppression of infection processes during the population
of the dark state also provides a natural explanation why the decay in
the quantum contact process is faster compared to directed
percolation, i.e., the exponent $\delta$ being larger. This behavior
is also consistent with the fact that introducing strong
dephasing, which couples dark states to bright states, drives the
transition towards the DP universality class \cite{Marcuzzi2015}.

For the quantum contact process, it is instructive to look at the dark
states of the continuous-time Hamiltonian for $L=2$. Then, the
Hamiltonian has two dark states, one being the trivial absorbing state
$\ket{00}$. However, the other dark state
$\ket{-}=(\ket{01}-\ket{10})/\sqrt{2}$ is non-trivial and maximally
entangled, arising from the destructive interference of the two terms
in the Hamiltonian. This state can be contrasted with the orthogonal
state $\ket{+}=(\ket{01}+\ket{10})/\sqrt{2}$, which is a superposition of
bright states. In the classical version of the contact process, there
are no superposition states, meaning the probability to find a
neighboring pair of cells $i,i+1$ in the $\ket{+}_{i,i+1}$ or
$\ket{-}_{i,i+1}$ state must be identical. This suggests to introduce
the quantity
\begin{equation}
  R(t) = \mathlarger{\sum\limits_i}\,\frac{\text{Tr}\{\ketbra{-}_{i,i+1}\rho(t)\}}{\text{Tr}\{\ketbra{+}_{i,i+1}\rho(t)\}},
\end{equation}
measuring the ratio between dark and bright states on neighboring
pairs to quantify the emergence of dark states and hence the
quantumness of the dynamics.

In Fig.~\ref{fig:dark}, we show the behavior of $R(t)$ both in the
continuous-time QCP regime and the DP regime ($\theta =
3\pi/4$). Strikingly, we find that $R(t)$ remains at a value
close to $1$ in the DP case, while it reaches almost $2$ in the QCP
regime. This clearly shows that the dark-to-bright ratio $R(t)$ can be
used to assess the difference between the universality
classes. Moreover, in the long-time limit, $R(t)$ approaches 2 almost
throughout the entire QCP region and only starts to decay when the
system is entering the anomalous DP region. This suggests that
also the intermediate regime is associated with destructive
interference effects from dark states.

\section{Methods}

\subsection{Quantum gates for coherent and dissipative dynamics}

Each discretized timestep consists
of (i) alternating layers of nearest-neighbor
controlled-rotation gates to implement the coherent infection dynamics and (ii) a layer of local probabilistic reset channels to realize the dissipative recovery process. The coherent gates are applied left/right with control qubits on the odd sublattice, followed by
left/right with control qubits on the even sublattice \cite{Chertkov2023}. Here, the two-site controlled rotation acting on control site $j$ and target site $k$ is given by
\begin{align}
CR^{(j\,k)}_{x,y}(\theta) =
\exp\!\left[-\frac{i\theta}{2}\, n_j \sigma_k^{x,y}\right].
\label{eq:cr_gate_main}
\end{align}

The dissipative step follows from a series of amplitude damping channels according to
\begin{equation}
  \mathcal{V}_{d,i}\rho=K_{i,0}(\phi)\,\rho\,K_{i,0}^\dagger(\phi)+K_{i,1}(\phi)\,\rho\,K_{i,1}^\dagger(\phi),
\end{equation}
with the Kraus operators $K_{i,\alpha}$ being given by
\begin{equation}
K_{0}(\phi)=
\begin{pmatrix}
1 & 0\\
0 & \cos(\phi/2)
\end{pmatrix},
\qquad
K_{1}(\phi)=
\begin{pmatrix}
0 & \sin(\phi/2)\\
0 & 0
\end{pmatrix}.
\label{eq:weakreset_kraus_main}
\end{equation}
For a gate-level implementation, we couple the system qubit $i$ to an ancilla $a$ initialized in $\ket{0}$ via a parametric swap or $R_{xy}$ gate
\begin{equation}
  R_{xy}(\phi) = \exp(-i \phi/4[\sigma^x_i \sigma^x_a + \sigma^y_i\sigma^y_a]).
\end{equation}
This produces an excitation probability of the ancilla of $p=\sin^2(\phi/2)$. The ancilla is measured and reset to $\ket{0}$. This measurement and reset operation causes quantum projection noise in the system, whose strength $\xi^2$ is given by the variance of the probability distribution. For the parametric reset process considered here, we obtain $\xi = \sin(\phi)/2$.

While the quantum projection noise only captures the fluctuations of
the diagonal elements of the density matrix, it is also instructive to
look at the off-diagonal coherences. Specifically, we have $\rho_{01} \to \sqrt{1-p}\rho_{01}$. For a situation, where strong reset gates with $\phi=\pi$ are probabilistically inserted into the circuit \cite{Chertkov2023}, the resulting channel damps the off-diagonal elements according to $\rho_{01} \to (1-p)\rho_{01}$, which is inherently stronger at small $p$.

\subsection{Matrix product state simulations}

We perform numerical simulations using a wave-function Monte-Carlo
algorithm \cite{Weimer2021}. Tensor operations are carried out using
the TenPy library \cite{Hauschild2024}. Continuous-time dynamics
is computed using a time-evolving block decimation algorithm.

\subsection{Estimation of the critical exponent}

We perform simulations up to a time $t = 25\,\tau/\theta $ and consider the
last third of the time-evolution of $\delta_\text{eff}(t)$ according to
Eq.~(\ref{eq:delta}) at $m=2$. The critical $\phi$ value is determined
as the point where the standard deviation of $\delta_\text{eff}(t)$ becomes
minimal.

\section{Acknowledgments}

  We thank M.~Darmst\"adter, E.~Gillman, and I.~Lesanovsky for
  fruitful discussions.

\bibliographystyle{nature}
\bibliography{../bib/bib}

\begin{thebibliography}{10}

\bibitem{Kasprzak2006}
Kasprzak, J. \emph{et~al.}
\newblock {Bose–Einstein} condensation of exciton polaritons.
\newblock \emph{Nature} \textbf{443}, 409--414 (2006).

\bibitem{Lai2007}
Lai, C.~W. \emph{et~al.}
\newblock Coherent zero-state and $\pi$-state in an exciton–polariton
  condensate array.
\newblock \emph{Nature} \textbf{450}, 529--532 (2007).

\bibitem{Syassen2008}
Syassen, N. \emph{et~al.}
\newblock Strong Dissipation Inhibits Losses and Induces Correlations in Cold
  Molecular Gases.
\newblock \emph{Science} \textbf{320}, 1329--1331 (2008).

\bibitem{Verstraete2009}
Verstraete, F., Wolf, M.~M. \& Ignacio~Cirac, J.
\newblock Quantum computation and quantum-state engineering driven by
  dissipation.
\newblock \emph{Nature Phys.} \textbf{5}, 633--636 (2009).

\bibitem{Baumann2010}
{Baumann}, K., {Guerlin}, C., {Brennecke}, F. \& {Esslinger}, T.
\newblock {Dicke quantum phase transition with a superfluid gas in an optical
  cavity}.
\newblock \emph{Nature} \textbf{464}, 1301--1306 (2010).

\bibitem{Carusotto2013}
Carusotto, I. \& Ciuti, C.
\newblock Quantum fluids of light.
\newblock \emph{Rev. Mod. Phys.} \textbf{85}, 299--366 (2013).

\bibitem{Lemeshko2013a}
{Lemeshko}, M. \& {Weimer}, H.
\newblock {Dissipative binding of atoms by non-conservative forces}.
\newblock \emph{Nature Commun.} \textbf{4}, 2230 (2013).

\bibitem{Marcuzzi2016}
Marcuzzi, M., Buchhold, M., Diehl, S. \& Lesanovsky, I.
\newblock Absorbing State Phase Transition with Competing Quantum and Classical
  Fluctuations.
\newblock \emph{Phys. Rev. Lett.} \textbf{116}, 245701 (2016).

\bibitem{Letscher2017}
Letscher, F., Thomas, O., Niederpr\"um, T., Fleischhauer, M. \& Ott, H.
\newblock Bistability Versus Metastability in Driven Dissipative Rydberg Gases.
\newblock \emph{Phys. Rev. X} \textbf{7}, 021020 (2017).

\bibitem{Helmrich2020}
{Helmrich}, S. \emph{et~al.}
\newblock {Signatures of self-organized criticality in an ultracold atomic
  gas}.
\newblock \emph{Nature} \textbf{577}, 481--486 (2020).

\bibitem{Wu2024}
{Wu}, X. \emph{et~al.}
\newblock {Dissipative time crystal in a strongly interacting Rydberg gas}.
\newblock \emph{Nature Phys.} \textbf{20}, 1389--1394 (2024).

\bibitem{Mi2024}
Mi, X. \emph{et~al.}
\newblock Stable quantum-correlated many-body states through engineered
  dissipation.
\newblock \emph{Science} \textbf{383}, 1332--1337 (2024).

\bibitem{Wu2024b}
Wu, L.-N. \emph{et~al.}
\newblock Indication of critical scaling in time during the relaxation of an
  open quantum system.
\newblock \emph{Nature Commun.} \textbf{15}, 1714 (2024).

\bibitem{Buchhold2017}
Buchhold, M., Everest, B., Marcuzzi, M., Lesanovsky, I. \& Diehl, S.
\newblock Nonequilibrium effective field theory for absorbing state phase
  transitions in driven open quantum spin systems.
\newblock \emph{Phys. Rev. B} \textbf{95}, 014308 (2017).

\bibitem{Carollo2019}
Carollo, F., Gillman, E., Weimer, H. \& Lesanovsky, I.
\newblock Critical Behavior of the Quantum Contact Process in One Dimension.
\newblock \emph{Phys. Rev. Lett.} \textbf{123}, 100604 (2019).

\bibitem{Gillman2019}
Gillman, E., Carollo, F. \& Lesanovsky, I.
\newblock Numerical simulation of critical dissipative non-equilibrium quantum
  systems with an absorbing state.
\newblock \emph{New J. Phys.} \textbf{21}, 093064 (2019).

\bibitem{Jo2021}
Jo, M., Lee, J., Choi, K. \& Kahng, B.
\newblock Absorbing phase transition with a continuously varying exponent in a
  quantum contact process: A neural network approach.
\newblock \emph{Phys. Rev. Res.} \textbf{3}, 013238 (2021).

\bibitem{Chertkov2023}
{Chertkov}, E. \emph{et~al.}
\newblock Characterizing a non-equilibrium phase transition on a quantum
  computer.
\newblock \emph{Nature Phys.} \textbf{19}, 1799--1804 (2023).

\bibitem{Hinrichsen2000}
Hinrichsen, H.
\newblock Non-equlibrium critical phenomena and phase transitions into
  absorbing states.
\newblock \emph{Adv. Phys.} \textbf{49}, 815--958 (2000).

\bibitem{Sieberer2025}
Sieberer, L.~M., Buchhold, M., Marino, J. \& Diehl, S.
\newblock Universality in driven open quantum matter.
\newblock \emph{Rev. Mod. Phys.} \textbf{97}, 025004 (2025).

\bibitem{Weimer2023}
{Weimer}, H.
\newblock {Quantum simulation gets openly critical}.
\newblock \emph{Nature Phys.} \textbf{19}, 1753--1754 (2023).

\bibitem{Makki2024}
Makki, N., Lang, N. \& B\"uchler, H.~P.
\newblock Absorbing state phase transition with Clifford circuits.
\newblock \emph{Phys. Rev. Res.} \textbf{6}, 013278 (2024).

\bibitem{Brady2024}
Brady, D., Ohler, S., Otterbach, J. \& Fleischhauer, M.
\newblock Anomalous Directed Percolation on a Dynamic Network Using Rydberg
  Facilitation.
\newblock \emph{Phys. Rev. Lett.} \textbf{133}, 173401 (2024).

\bibitem{Vojta2005}
Vojta, T. \& Dickison, M.
\newblock Critical behavior and Griffiths effects in the disordered contact
  process.
\newblock \emph{Phys. Rev. E} \textbf{72}, 036126 (2005).

\bibitem{Wintermantel2021}
Wintermantel, T.~M. \emph{et~al.}
\newblock Epidemic growth and Griffiths effects on an emergent network of
  excited atoms.
\newblock \emph{Nature Communications} \textbf{12}, 103 (2021).

\bibitem{Brady2024b}
Brady, D. \emph{et~al.}
\newblock Griffiths phase in a facilitated Rydberg gas at low temperatures.
\newblock \emph{Phys. Rev. Res.} \textbf{6}, 013052 (2024).

\bibitem{Schiulaz2015}
Schiulaz, M. \& M{\"u}ller, M.
\newblock Ideal quantum glass transitions: Many-body localization without
  quenched disorder.
\newblock \emph{AIP Conf. Proc.} \textbf{1610}, 11--23 (2014).

\bibitem{Fleischhauer2005}
Fleischhauer, M., Imamoglu, A. \& Marangos, J.~P.
\newblock Electromagnetically induced transparency: Optics in coherent media.
\newblock \emph{Rev. Mod. Phys.} \textbf{77}, 633--673 (2005).

\bibitem{Diehl2008}
Diehl, S. \emph{et~al.}
\newblock Quantum states and phases in driven open quantum systems with cold
  atoms.
\newblock \emph{Nature Physics} \textbf{4}, 878--883 (2008).

\bibitem{Kraus2008}
Kraus, B. \emph{et~al.}
\newblock Preparation of entangled states by quantum Markov processes.
\newblock \emph{Phys. Rev. A} \textbf{78}, 042307 (2008).

\bibitem{Weimer2010}
{Weimer}, H., {M{\"u}ller}, M., {Lesanovsky}, I., {Zoller}, P. \&
  {B{\"u}chler}, H.~P.
\newblock A Rydberg quantum simulator.
\newblock \emph{Nature Phys.} \textbf{6}, 382--388 (2010).

\bibitem{Marcuzzi2015}
Marcuzzi, M. \emph{et~al.}
\newblock Non-equilibrium universality in the dynamics of dissipative cold
  atomic gases.
\newblock \emph{New Journal of Physics} \textbf{17}, 072003 (2015).

\bibitem{Weimer2021}
Weimer, H., Kshetrimayum, A. \& Or\'us, R.
\newblock Simulation methods for open quantum many-body systems.
\newblock \emph{Rev. Mod. Phys.} \textbf{93}, 015008 (2021).

\bibitem{Hauschild2024}
Hauschild, J. \emph{et~al.}
\newblock Tensor network Python (TeNPy) version 1.
\newblock \emph{SciPost Phys. Codebases} 41 (2024).

\end{thebibliography}

\end{document}